# Optimizing Cost, Delay, Packet Loss and Network Load in AODV Routing Protocol

Ashutosh Lanjewar
M.Tech (DC) Student
T.I.E.I.T. (TRUBA)
Bhopal (M.P), India
ashutoshlanjewar39@gmail.com

Neelesh Gupta
Department of Electronics &Communication
T.I.E.I.T. (TRUBA)
Bhopal (M.P), India
neelesh.9826@gmail.com

***Abstract*: AODV is Ad-hoc On-Demand Distance Vector. A mobile ad-hoc network is a self-configuring network of mobile devices connected by wireless. MANET does not have any fixed infrastructure. The device in a MANET is free to move in any direction and will form the connection as per the requirement of the network. Due to changing topology maintenance of factors like Packet loss, End to End Delay, Number of hops, delivery ratio and controlling the network load is of great challenge. This paper mainly concentrates on reducing the factors such as cost, End-to-End Delay, Network Load and Packet loss in AODV routing protocol. The NS-2 is used for the simulation purpose.**

## I. INTRODUCTION

Mobile Ad-Hoc network mainly concentrates on wireless communication without any fixed infrastructure. Wireless communication has wide application in Security zones .In past there is only a fixed wireless communication network exists where communication range is bonded. Now there advanced Ad-Hoc network and Mobile Ad-Hoc network are introduced where all nodes share data among themselves. The nodes in AODV may connect and leave the network at any time [10].All Ad-Hoc routing protocol have different routing strategies so factors such as End to End Delay, Traffic Overhead and packet delivery ratio and power consumption gets vary .Routing mainly deals with the route discovery between the source and destination [4].Nodes in network change the position as per requirement of system so topology varies time to time. The routing Protocols are mainly divided in to Routing and Reactive Protocol. Proactive routing protocols (e.g.OLSR) are table-driven. Link-state algorithms maintain a full or partial copy of the network topology and costs for all known links. The reactive routing protocols (e.g. AODV) create and maintain routes only if these are needed, on demand. They usually use distance-vector routing algorithms that keep only information about next hops to adjacent neighbors and costs for paths to all known destinations. Thus, link-state routing algorithms are more reliable, less bandwidth-intensive, but also more complex and compute and memory-intensive. AODV routing protocol is a reactive routing protocol. AODV is a related to the Bellman-Ford distant vector algorithm. In AODV a route to a destination is determined when a node wants to send a packet to that destination. Routes are maintained as long as they are needed by the source. When the packet is transmitted from source to destination there are many nodes involved between the successful receptions of packets. ADOV routing protocol uses RouteRequest (RREQ) RouteReply (RREP) and RouteError (RERR) as a control signal. When a source node desires to send a message to some destination node and does not have a valid route to that destination it looks for a Path to locate the other node. Source node sends a RREQ packet to its neighbors, which then forward the request to their neighbors, and the process go on until route to the destination is located [2]. During the process of forwarding the RREQ, the entry of intermediate nodes get record in to their routing tables which include the address of the neighbors from which the first copy of the broadcast packet is received. This will help to find a path. If in case some additional copies of the same RREQ are received later than these packets are discarded. Once the RREQ reaches the destination node, the destination or intermediate node responds by sending a RREP packet back to the neighbor from which it first received the RREQ. When packet transmission is in progress various factors play measure role .It is observed that packet may get drop in between due to bad linkage quality and lack of proper communication channel between the nodes. Sometimes communication gets successful but the backend factors such as End to End delay, Power consumption, Routing overhead and hop limit really makes the network really costly and unreliable one. In AODV the routing table plays the important role. The route table includes the entry at each node with the information regarding the sequence number for IP address of destination node. The RREQ, RREP and RERR commands are received by node utilized for the updating of the sequence number. The destination node can increment its sequence number when there is time for source node to start a route search or when there is time for destination node to generate the RREP message against the RREQ response of source node. In routing table the route gets updated with new sequence numbers when it is higher than the destination sequence numbers. There are other two possibilities, the first one is when the new sequence number and destination sequence numbers are equal but if sum number of hop plus one additional one hop in new sequence routing table is smaller than hop count in the existing destination sequence number and secondly when the existing sequence number is unknown.

The rest of this paper is ordered as follows. The related works are discussed in Section II, Section III represents working of AODV routing protocol and Section IV gives idea regarding the proposed work. Section V gives detail of simulation results and its discussion. Section VI provides conclusion and future work whereas section VII represents References.

## II. RELATED WORK

AODV is reactive routing protocol. It is simple, efficient and effective routing protocol having wide application [14]. The topology of the network in AODV gets change time to time so dealing with same and as well as maintaining the Cost, End-to-End, Network Load and Packet Loss is great challenge. Various researches have been carried out on above factors.Lalet.al. [13] implemented new NDMP-AODV that is able to provide low end-to-end delay and high packet delivery ratio, while keeping low routing overhead. In future work they improve the route selection process of NDMP-AODV so that it can select routes that can satisfy user application requirements. Raj Kumar G.et.al [15] evaluated the AODV and DSR on parameter such as Throughput, Delay, Network Load and Packets Drop against pause time .They observed that AODV performs well in the presence of noise gives better throughput level with less delay, consumes less energy and less packets get drop .Maurya1et.al. [2] Compared on-demand routing protocols that is reactive and proactive routing. They observed that reactive protocol offers quick adaptation to mobile networks with low processing and low bandwidth utilization. In [3] Das et.al. two on-demand routing protocols, DSR and AODV had been compared. In future, they have studied more routing protocols such as DSDV, TORA based on parameters such as fraction of packet delivery, end to end delay and routing overhead.Yanget.al. [5] compared the AODV, R-AODV and SR-AODV .From simulation they have concluded that SR-AODV improves the performance of AODV in most metrics, as the packet delivery ratio, end to end delay, and Power consumption.Yanget.al.[7] analyzed the performances of AODV and M-AODV they observed that in M-AODV route discovery succeeds in fewer tries than AODV. When the simulation is carried out they conclude that M-AODV improves the performance of AODV in most metrics, as the packet delivery ratio, end to end delay, and energy consumption .Li et.al. [6] evaluated the TRP with S–AODV and it is observed that TRP improves network performance in terms of energy efficiency and average routing delay. In [4] Thanthryet.al.they verified the EMAODV with the AODV. The results obtained from the simulations show that EMAODV performs better than AODV in terms of throughput, number of route discoveries, control overhead and packet drops but, the average end-to-end delay of EM-AODV was found to be higher than AODV.Khelifaet.al.[1] investigated the performances of M-AODV and AODV they observed route discovery succeeds in that M-AODV improves the performance of AODV in terms of metrics, packet delivery ratio, end to end delay, and energy consumption. In future they studied the implementation of Energy AODV mechanism to conserve more energy. Sharma et al.[8] evaluated the effect of different scheduling algorithms for AODV and modified AODV. They reduce the average delay between the nodes communication. Wei et.al [9] worked on Demand Distance Vector (IPODV) routing protocol considering the topological feature of the power-line network. In future they work on the routing maintenance mechanism and the neighbor table management of the AODV routing Protocol. Chaurasia et.al. examined[11] on OLSR, DSDV, DSR, AODV, and TORA protocols They observed that due to the infrastructure less structure of protocol security and power awareness is difficult to achieve in mobile ad hoc networks .In future they work on core issues of security and power consumption in these routing protocol.M.Ushaet.al. [12] implemented new advanced AODV name RE-AODV (Route-Enhanced AODV). They observed routing overhead is reduced by 25% and end to end delay of packets 11% as compared to normal AODV protocol. It has been observed in AODV routing protocol that power consumption is more which make AODV a costly one .The end-to-end delay is more, there increase the chances for loss of information while transaction between the source node and destination node. So the effort are required to be taken regarding the reduction of power consumption and end-to-end delay in order to reduce the costing in implementation of AODV routing protocol.

The related work in the field of AODV routing protocol really creates the motivating impact on the mind for further research .The implementation of the AODV routing protocol with all features such as less end-to-end delay, maintenance of network Load, Packet loss and cost is really a challenging one. The proposed work mainly concentrates on implementation of all above parameters. This implementation will really prove advantageous for the networking technology.

## III. AODV ROUTING PROTOCOL

AODV is a self-starting and dynamic algorithm where the large number of nodes can participate for establishing communication and maintaining AODV network. The topology of AODV changes time to time as the nodes are not fixed to any standard position. In AODV hello messages are used to detect and monitor links between the nodes. An active node periodically broadcasts a Hello message to all its neighboring nodes. If in case the nodes fail to transmit hello message to neighboring node, the complete network will collapse due to link breakage. AODV uses mainly three message types Route Requests (RREQs), Route Replies (RREPs) and Route Errors (RERRs).These message are carried through UDP and IP headers. When the source node want to send data to the destination node it send the RREQ message .This RREQ message may be received directly by the destination node or intermediate node. In AODV the destination sequence number is generated. During the period when the node request for the route discovery it is provided with destination sequence numbers. A requesting node is requiring to select the one with greatest sequence number. Then the route is made available by unicasting a RREP back to the source node from RREQ is send. AODV mainly deals with route table. In route table the information of all the transaction between the nodes are kept. The routing request has following sections Source address, Request ID, Source sequence number, destination address, destination sequence number and hop count. The route request Id gets incremented during single transaction from source node. At the destination node the Request ID and source address are verified. The route request with same request ID is discarded and no route reply message will generate. Every route request has its TTL i.e. Time To Live and during this time period the route request can be retransmitted if reply is not received from destination node. If the route is valid than destination node unicast the route reply message to the source node. The route Reply has following sections source address, destination address, destination sequence number, hop count and life time. Hop count defines number of nodes utilized for data. When node involve in active transaction gets lost, a route error (RERR).The message format of route request, route reply and route error are given below.

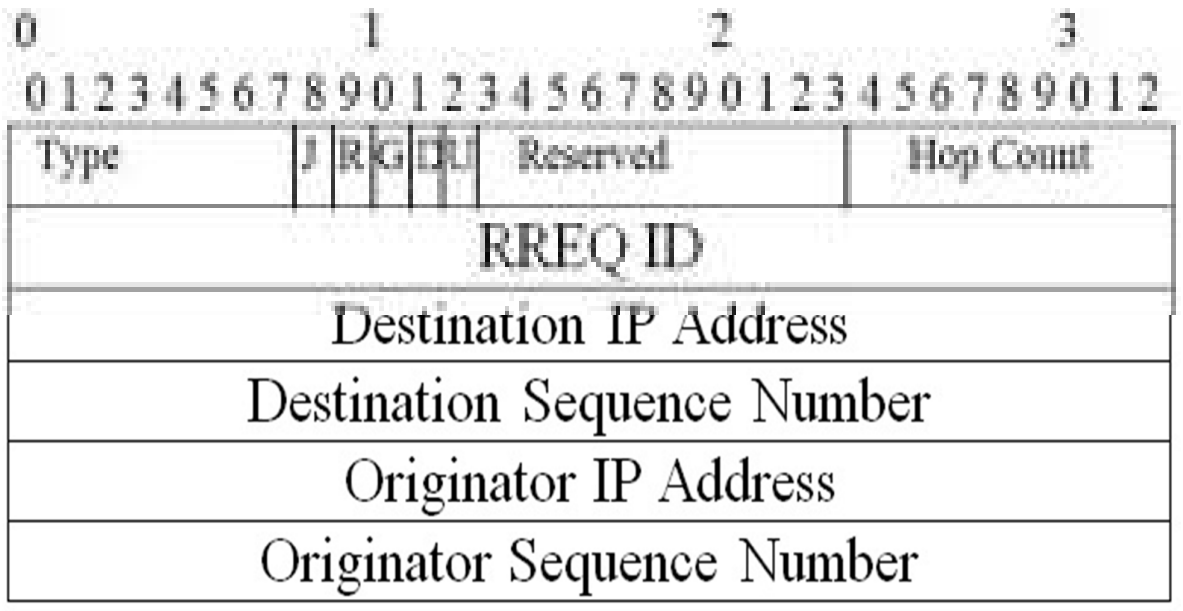

**Figure 1. Message Format of (RERQ)**

In figure 1Type of RREQ is 1.J represents the Join flag and R represents Repair flag both are reserved for multicasting purpose. G represents Gratuitous RREP flag which indicate that data is unicast to the node with specified Destination IP address field. D represents that only destination will respond to the RREQ and no intermediate node will act. U represents that sequence Number is unknown.

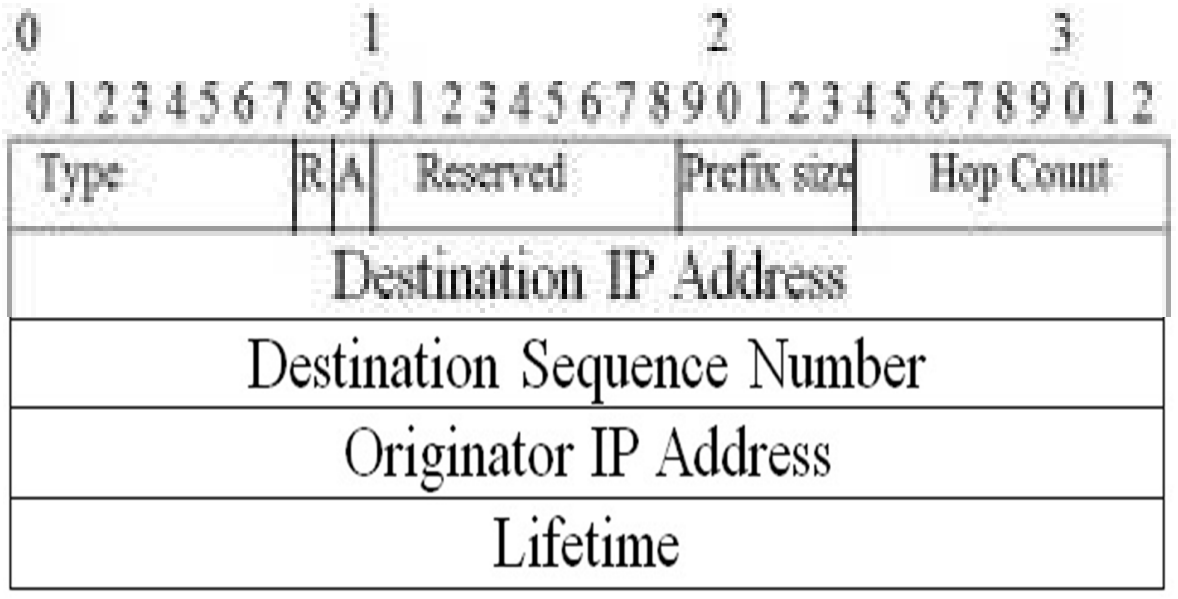

**Figure 2. Message Format of (RREP)**

In figure 2 Type of RREP is 2. R represents Repair flag and it is used for multicast. A represents Acknowledgment required and Reserved is indicated by 1 when network is ready to give route reply or by 0 then no reply will be given to route request. Prefix size represents that next hop may be used for any nodes with the same routing prefix**.**
Now in figure 1 and figure 2 Hop count represents the number of hops required during the retransmissions. Destination IP Address represents IP address of destination to which route is to be generated. Destination Sequence Number is always related with the route. Originator IP Address represents the source from which the RREQ is generated whereas; the Life time is the time period during which the node receives the RREP to validate the route.

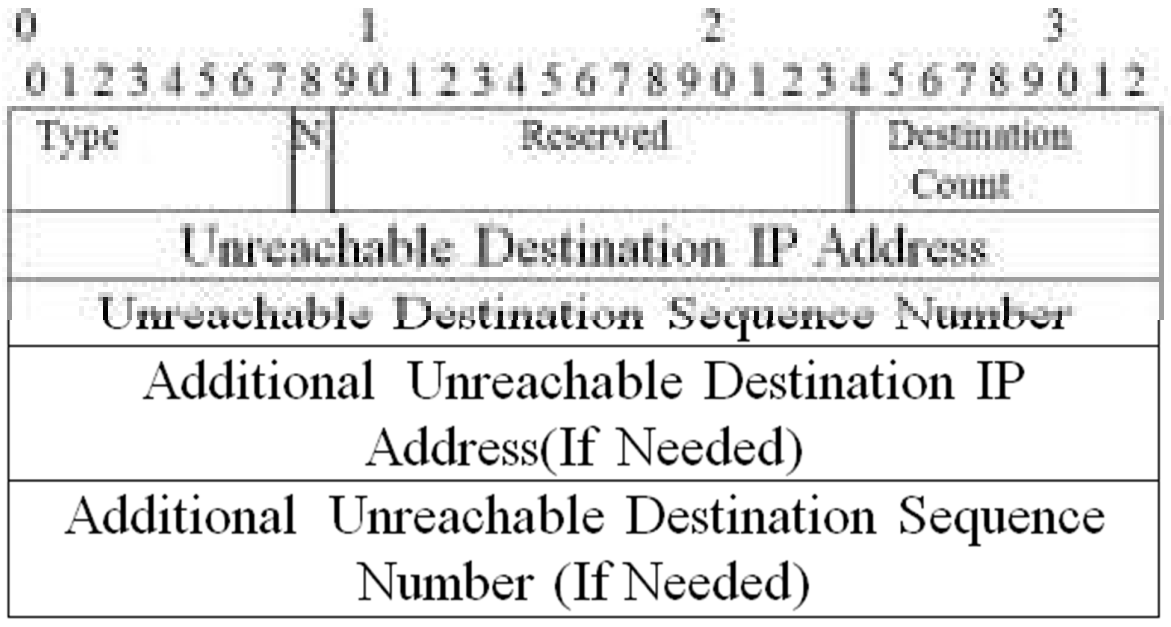

**Figure 3. Message Format of (RRER)**

In figure 3 Type of RRER is 3. N represents that flag will not get delete. Reserved is sent as 0 represents that RERR is ignored. Destination Count represents the number of destinations that are out of reach and this count will included in the message. Unreachable Destination IP Address represents the IP address of destination is not reachable due problem in link whereas Unreachable Destination Sequence Number represents sequence number of destination whose IP address is not reachable due to link breakage.

## VI.PROPOSED METHOD

The performance comparison of Normal AODV and newly generated AODV routing protocols are analyzed and tested for 40 nodes when simulations are carried on NS-2 simulator. The AODV routing protocol will perform better than past ones. The cost and end-to-end delay will get reduce also there by minimize the network load and packet loss. Special concentration is given on controlling the hop limit. The number of nodes utilized for single transaction from assigned source to destination will get reduced. As hop limit is achieved indirectly it affects network load, end-to-end delay and indirectly the probability of packet loss. The ultimate cost of the network gets reduce in AODV routing protocol. In the project the Euclidean distance between the nodes is calculated which gives the idea regarding time require to transfer data from source to destination and distance between the source and destination. Thus the Euclidean distance formula is used for determining the costing of the network. The AODV network with nodes P, Q, R, S, and T is given in figure 4. Consider the two dimension Euclidean space.

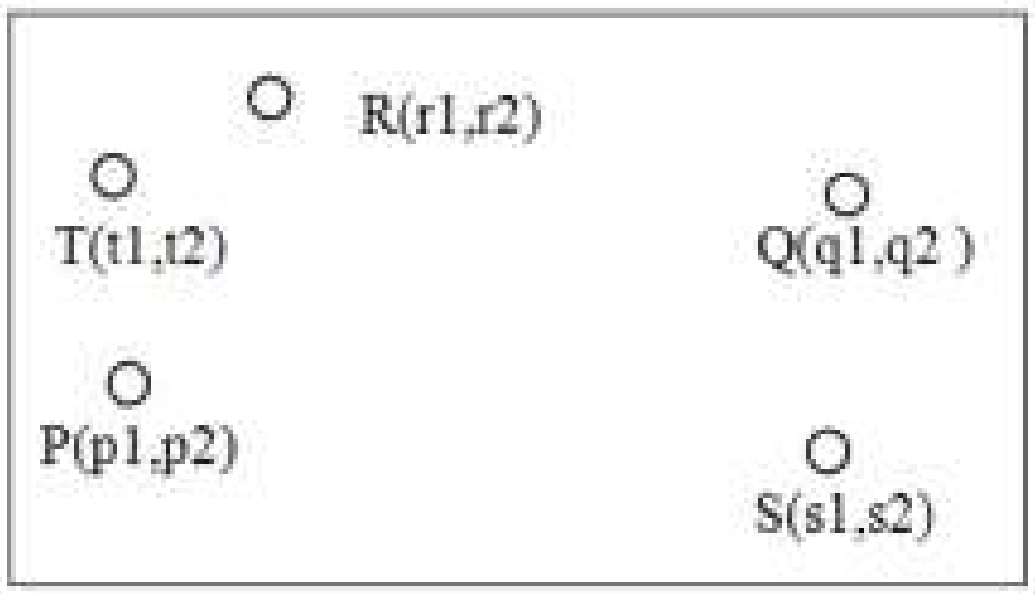

**Figure 4. Nodes in two dimension Euclidean space**

. In order to find the Euclidean distance between two nodes P and Q , first of all P and Q are described with coordinates (p1,p2) and (q1,q2) respectively . In first step length between the P and Q is given by |p1 - q1| and |p2 - q2|.Secondly the Pythagorean Theorem is between the two length gives ((p1 - q1) ^2 + (p2 - q2) ^2) ^ (1/2). So the distance between two points P = (p1, p2) and Q = (q1, q2) in two dimensional space is there given asx = $\sqrt{(p1-q1)^2+(p2-q2)^2}$ .Similarly the distance between two points P = (p1, p2, ..., pn) and Q = (q1, q2, ..., qn) in n dimensionsEuclidean space can be given as can be given as$\sqrt{(p1-q1)^2+(p2-q2)^2+\cdots+(pn-qn)^2}$.

The key advantages of the proposed work are multiple. The good network mainly concerns with the efficient transfer of data, minimum costing, less packet loss and Network Load. The performance of Normal AODV and AODV routing protocols are compared based on the performance metrics which are given below. The four parameter are evaluated against number of transfers.

*Cost*: It depends on number of nodes utilized, power consumed and packet loss.

*End to End delay*: It is the difference between the packets received time and packet sent time.
*Packets drop*: It is the number of packets lost in transit.

*Network Load*: The total traffic (bits/sec) received by the network layer from the higher MAC that is accepted and queued for transmission.

## V. SIMULATION RESULTS AND DISCUSSION

The simulation has been done for 40 nodes using Network Simulator 2.35 in an area of size 1000 m x 1000m. The performance metrics such as cost, end to end delay and Network Load are evaluated against number of transfers for both Normal AODV and New advance AODV Routing protocols and are shown below. The red colour curve represents the Normal AODV protocol while the green colour curve represents the proposed new advance AODV protocol. The Simulation Parameters are given below

| | |
|---|---|
| Number of Nodes | 40 |
| Routing Protocol | AODV |
| Traffic Source | CBR |
| Area | 1000 m x 1000 m |
| Mac Type | IEEE 802. 11 |
| Tool | NS-2.35 |

**Table I –Simulation Parameters**

In Figure 5. Number of Data transfers is plotted against the cost. In the graph only three data transfers are consider .It is observed that cost require in a new advance AODV routing is very less as compare with normal AODV. Cost in Proposed AODV simulation touches the lower level of 153 units.

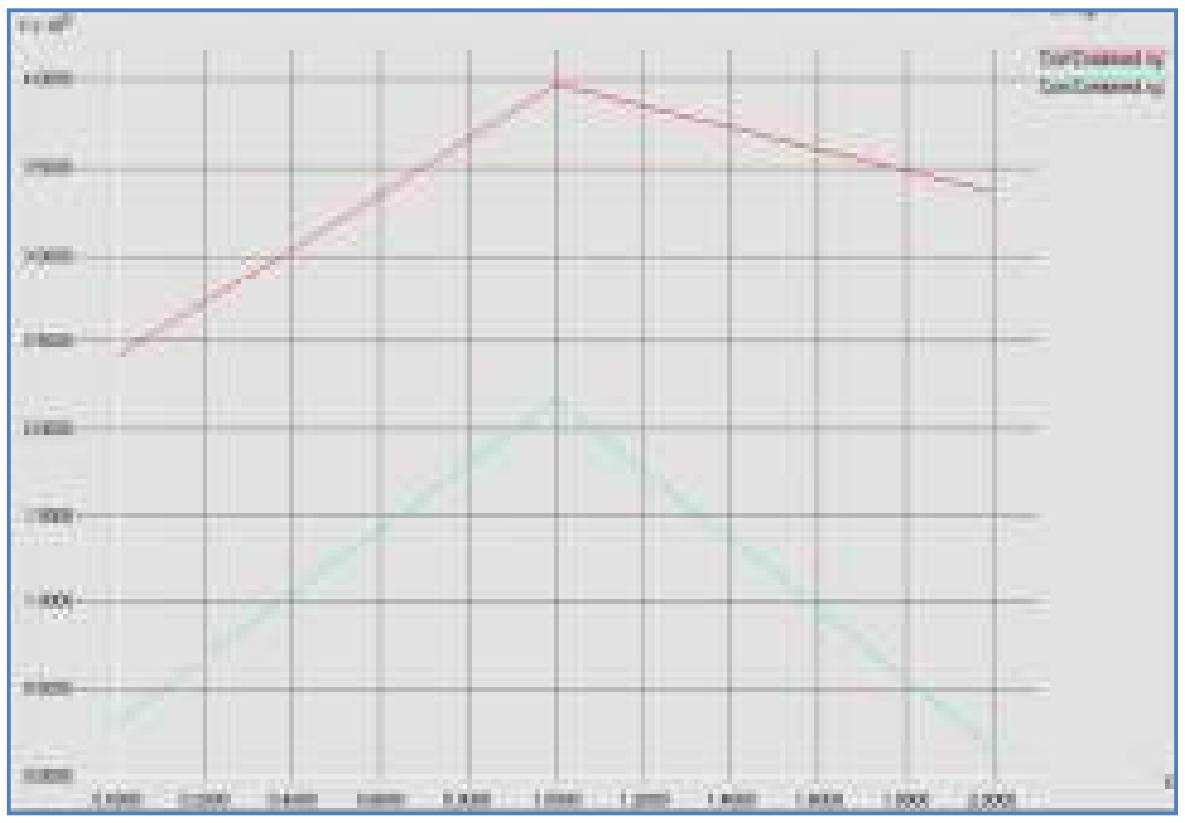

**Figure5.Number of Data Transfers versus Cost**

In figure .6 the Number of data transfers is plotted against delay. It is observed from graph that Proposed AODV has lowest delay in all data transfers as compare to normal AODV routing protocol.

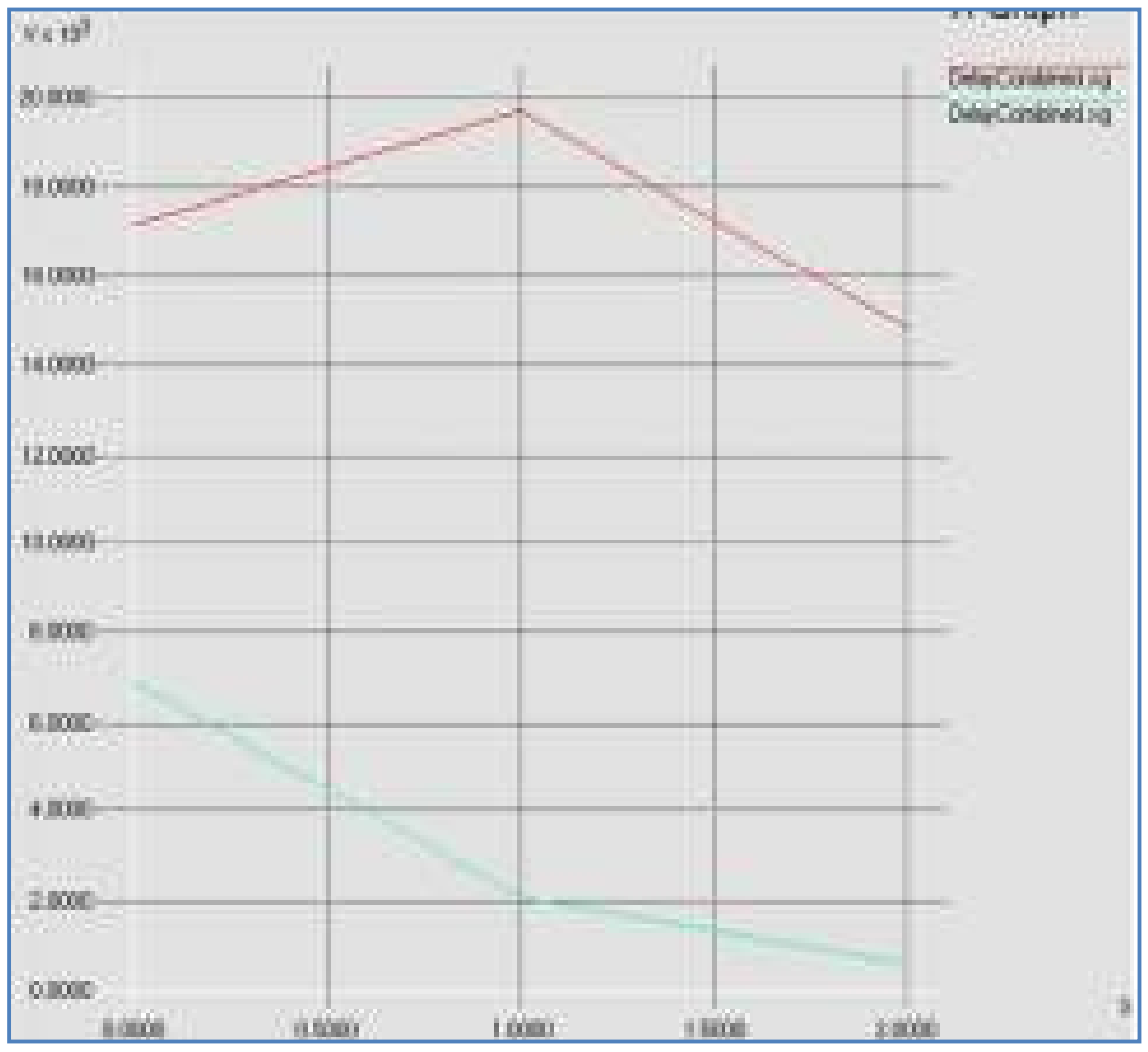

**Figure 6. Number of Data Transfers versus Delay (ms)**

In figure 7. The Number of data transfers is plotted against Packet loss. It is observed from graph that Proposed AODV has low packet loss as compare with normal AODV routing Protocol.

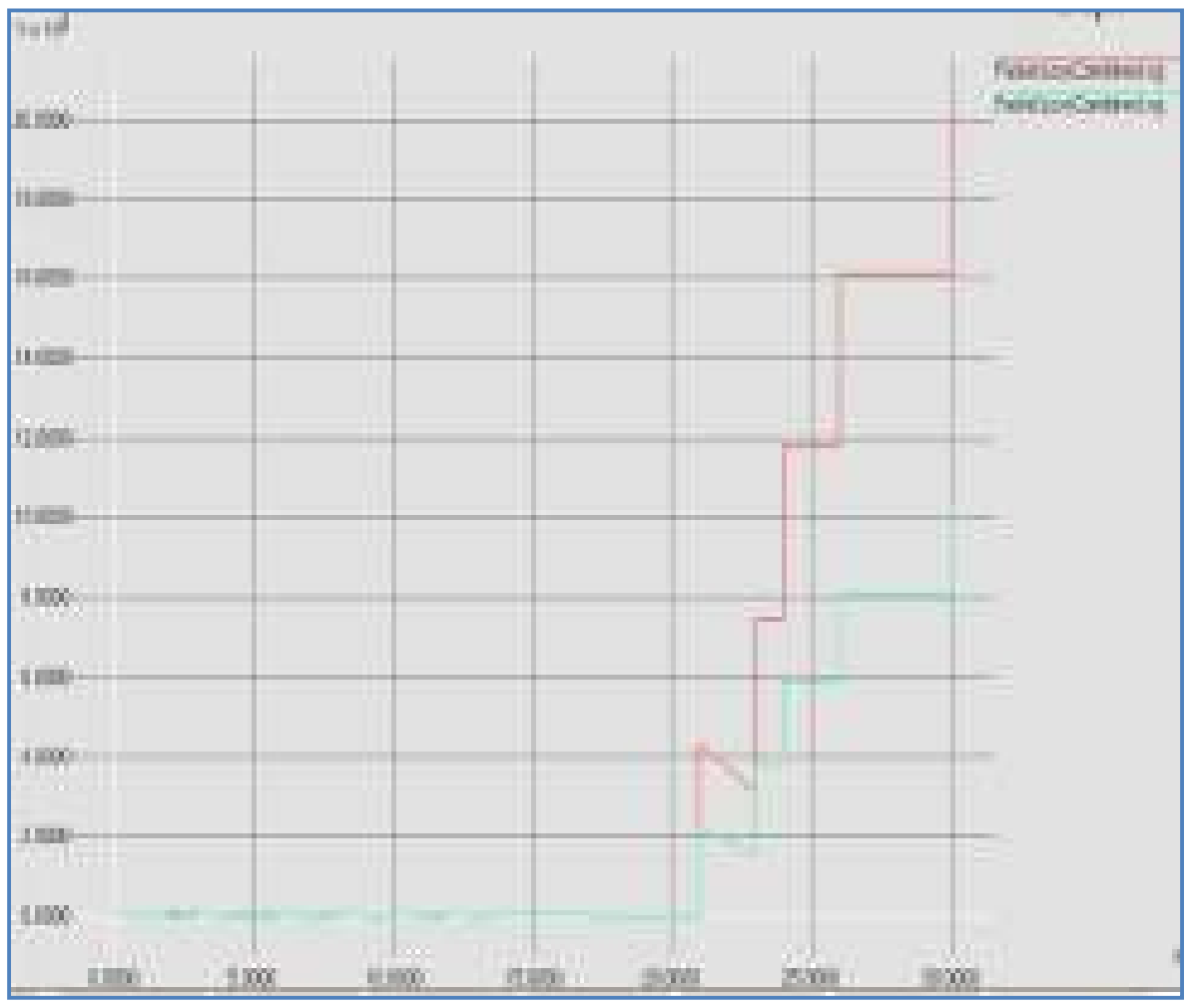

**Figure 7. Number of Data Transfers versus Packet Loss**

In figure 8 the Number of data transfers is plotted against Network Load. It is observed from graph that Proposed AODV has negligible network load in all data transfers as compare to normal AODV routing protocol.

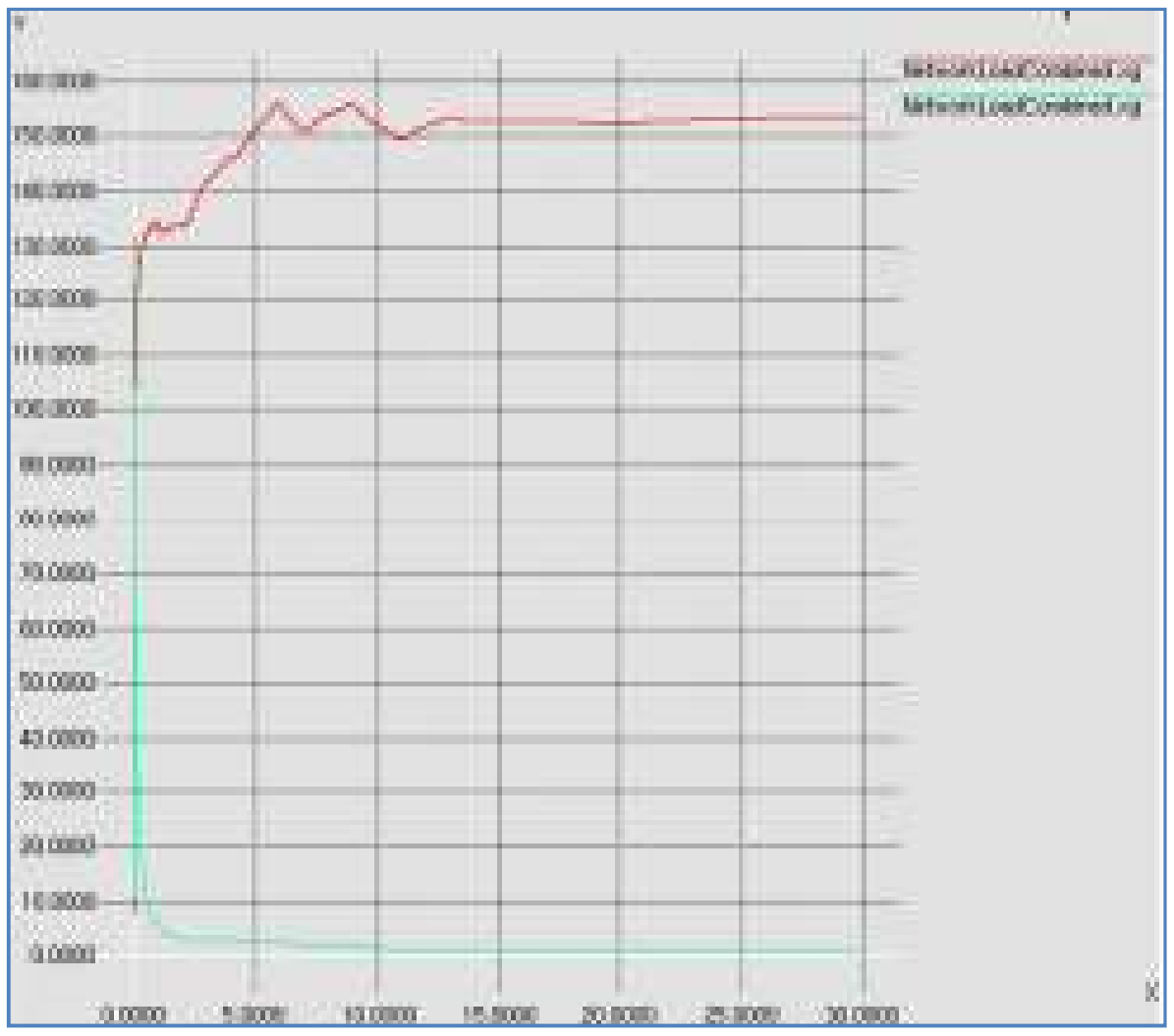

**Figure8.Number of Data Transfers versus Network Load**

## VI. CONCLUSION AND FUTURE WORK

The performance metrics such as Cost, Delay, Network Load and Packets Drop are evaluated against Number of transfers for both Normal AODV and new advance AODV with number of mobile nodes of up to 40 using NS-2.35. As the number of nodes is increased, still new advance AODV performs well and yields better throughput level with less delay and consumes less energy. Despite having high Network load new advance AODV is able achieve less packets Drop when compared to Normal AODV protocol. In this simulation new AODV has the all-round performance.

.

## V.REFERENCES

[1]Khelifa S., Maaza Z.M., “An Energy Multi-path AODV Routing Protocol in Ad Hoc Mobile Networks” IEEE International Symposium on Communications and Mobile Network , 2010 Conference Publications, pp.1-4, 2010.

[2]Maurya1 P.K., Sharma G., Sahu V., Roberts A. and Srivastava M., “An overview of AODV Routing Protocol” International Journal of Modern Engineering Research (IJMER), Vol.2, Issue3, pp.728-732, 2012.

[3] Das S.R., Perkins C.E., Royer E.M., “Performance Comparison of Two on-demand Routing Protocols for Ad-Hoc Networks”, 19th annual joint conference of the IEEE Computer and communication Societies, IEEE Procc., pp.3-12, Vol.-1, Isreal, INFOCOM, 2000.

[4]Thanthry N, Kaki S. R., Pendse R., “EM-AODV: metric based enhancement to aodv routing protocol”, IEEE 64th Vehicular Technology Conference, pp.1-5, 2006.

[5]Yang H. , Li Z., "A Stability Routing Protocols base on Reverse AODV", IEEE International Conference on Computer Science and Network Technology, Vol.4, pp.2419-2423, 2011.

[6]Li L., Chigan C., "Token Routing: A Power Efficient Method for Securing AODV Routing Protocol", IEEE International Conference on Networking, Sensing and Control, pp.29-34, 2006.

[7]Yang H., Li Z., "Simulation and Analysis of a Modified AODV Routing Protocols", IEEE International Conference on Computer Science and Network Technology, Vol.3, pp.1440-1444, 2011.

[8]Sharma D.K., Kumar C., Jain S., Tyagi N., "An Enhancement of AODV Routing Protocol for Wireless AdHoc Networks", IEEE International conference on Recent Advances in Information Technology , pp-290-294, 2012.

[9]Wei G., Jin W., Li H., "An Improved Routing Protocol for Power-line Network based on AODV" IEEE International Conference on Communications and Information Technologies, pp.233-237, 2011.

[10]Gupta N, Gupta R., "Routing Protocols in Mobile Ad-Hoc Networks: an Overview", IEEE International Conference on Emerging Trends in Robotics and Communication, pp.173-177, 2010.

[11]Chaurasia N., Sharma S.,Soni D., "Review Study of Routing Protocols and Versatile challenges of MANET"IJCTEEVolume2, Issue 1, pp.150-157, 2012.

[12] M.Usha, S.Jayabharathi, Banu R.S., "RE-AODV: An Enhanced Routing Algorithm for QoS Support in Wireless Ad-Hoc Sensor Networks" IEEE International conference on Recent Trends in Information Technology, pp.567-571, 2011.

[13]Lal C., Laxmi V. and Gaur M.S., "A Node-Disjoint Multipath Routing Method based on AODV protocol for MANETs", IEEE 26th International Conference on Advanced Information Networking and Applications (AINA), pp.399-405, 2012.

[14]Perkins C.E, Royer E., "Ad-Hoc On-Demand Distance Vector Routing", IEEE Workshop on Mobile Computing Systems and Applications, pp.90-100, 1999.

[15]Rajkumar G., Kasiram R. and Parthiban D "Optimizing Throughput with Reduction in Power Consumption and Performance Comparison of DSR and AODV Routing Protocols", International Conference on Computing, Electronics and Electrical Technologies, pp.943-947, 2012.

AUTHORS PROFILE

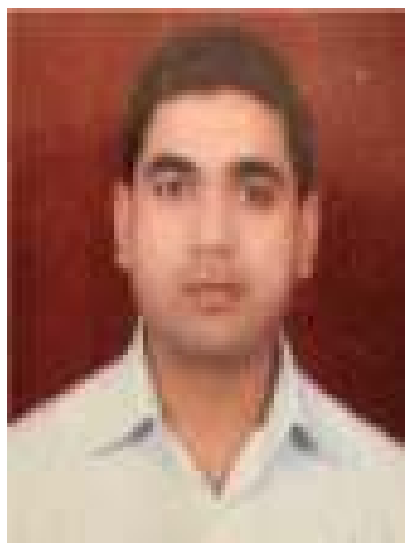

Ashutosh Lanjewar is Pursuing M.Tech in Digital Communication from Truba Institute of Engineering and Information Technology (T.I.E.I.T.), Rajiv Gandhi Proudyogiki Vishwavidyalaya (RGPV), Bhopal (M.P.) - India. He has completed his B.E. in Electronics and Telecommunication in 2007 from G.H.Raisoni College of Engineering, Nagpur (Maharashtra) - India. His area of research Interest is Wireless Communication and Networking.

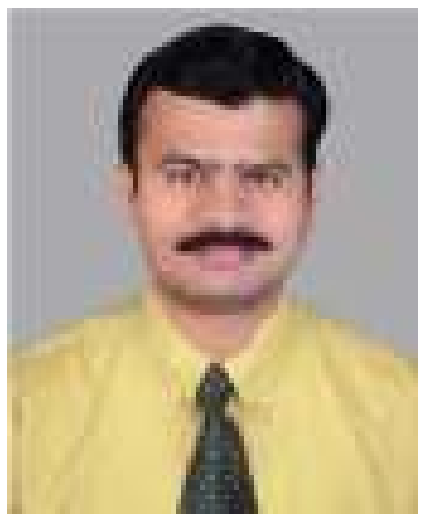

Neelesh Gupta is Pursuing Ph.D in Electronics and Communication from Rajiv Gandhi Technical University (RGTU), Bhopal (M.P.)-India. He has a rich experience of teaching in various Technical institutions of reputed in MP-India. He is having more than 10 years of teaching Experience. Presently he is an Assistant Professor in Truba Institute of Engineering and Information Technology (T.I.E.I.T.), Bhopal (M.P.) - India. He has earned his M.Tech degree in Microwave and Millimeter Wave in 2007 from MANIT, Bhopal. His area of research Interests are Wireless Communication, Microwaves and Digital Signal Processing. He has presented a number of research papers in various National, International conferences and reputed International Journals. He is Life time member of IETE, New Delhi.